\documentclass[9pt]{article}
\usepackage{mathrsfs}
\usepackage{amssymb}
\usepackage{amsfonts}
\usepackage{latexsym}
\usepackage{graphicx}
\usepackage{amsfonts,amssymb}

\begin{document}

\title{\textbf{Enhance the Efficiency of Heuristic Algorithm for Maximizing Modularity $Q$}}
\author{Yanqing Hu$^{1}$, Jinshan Wu$^{2}$, Zengru Di$^{1}$\footnote{Author for correspondence: zdi@bnu.edu.cn}\\
\\\emph{ 1. Department of Systems Science, School of Management,}\\
\emph{Beijing Normal University, Beijing 100875, P.R.China}
\\\emph{2. Department of Physics \& Astronomy, University of British Columbia, }\\ \emph{Vancouver, B.C. Canada, V6T 1Z1}}
\maketitle

\begin{abstract}
Modularity $Q$ is an important function for identifying community
structure in complex networks. In this paper, we prove that the
modularity maximization problem $\max_{S\in
\mathbb{S}}\,\,\bar{Q}=\emph{Tr}(S^{T}BS)$ is equivalent to a
nonconvex quadratic programming problem $\max_{S\in
\mathbb{S}}\,\,Q_{m}=\emph{Tr}(S^{T}(B+D)^{m}S)$. This result
provide us a simple way to improve the efficiency of heuristic
algorithms for maximizing modularity $Q$. Many numerical results
demonstrate that it is very effective.
\end{abstract}



\section {Introduction}
Complex network has received an enormous amount of attention in
recent years \cite{1,Newman reviwe,new reviwe}. Scientists have
become interested in the study of networks describing the topologies
of wide variety of systems such as the world wide web, social and
communication networks, biochemical networks and many more. Based on
complex networks many quantitative methods can be applied so as to
extract the characteristics embedded in the system. One of the
important quantitative methods is to analysis the community
structure \cite{1,Newman reviwe,new reviwe}. Distinct communities
within networks can loosely be defined as subsets of nodes which are
more densely linked, when compared to the rest of the network. Nodes
belonging to a tight-knit community are more than likely to have
other properties in common. In the world wide web, community
analysis has uncovered thematic clusters. In biochemical or neural
networks, communities may be functional groups \cite{1,3,GN}, and
separating the network into such groups could simplify the
functional analysis considerably. As a result, the problem of
identification of communities has been the focus of many recent
efforts.

Maximizing modularity $Q$ is the most widely accepted method for
detecting community structure among many algorithms
\cite{3,GN,4,6,9,10,12,14,17,18,20,Newman fast algorithm,Newman
mixture algorithm,Systematic,Newman evaluating algorithm,2005 pans
community defination,Lshell,EO,Potts}, although modularity index has
been proved that it may fail to identify small modules \cite{23}.
Modularity $Q$ was presented as a index of community structure by
Newman and Grive, which was introduced as
$Q=\sum_{r}{(e_{rr}-a_{r}^{2})}\label{Q}$, where $e_{rr}$ are the
fraction of links that connect two nodes inside the community $r$,
$a_{r}$ the fraction of links that have on or both vertices in side
the community $r$, and sum extends to all communities $r$ in a given
network. Note that this index provides a way to determine if a
certain description of the graph in terms of communities is more or
less accurate. Generally speaking, the larger the value of $Q$, the
more accurate is a partition into communities. So maximizing
modularity $Q$ can detect community structures. There are many
algorithms of maximizing $Q$ directly such as extremal optimization
(EO) \cite{EO}, greedy algorithm \cite{10} and other optimal
algorithm. In fact, they are usually heuristic algorithms for
modularity maximization problem and this problem has been proved to
be a NPC in the strong sense by Ulrik Brandes \emph{et al} \cite{Q
NPC}.

Can we improve the efficiency of corresponding heuristic algorithms
by detailed investigation of mathematic structure of modularity $Q$?
According to ref \cite{17}, $\max Q=\ \sum_{r}{(e_{rr}-a_{r}^{2})}$
can be simplified as $\max \bar{Q}=\emph{Tr}(S^{T}BS)$. In this
paper, we proved that $\max \bar{Q}=\emph{Tr}(S^{T}BS)$ is a
nonconvex quadratic $0-1$ programming. Assume
$D=diag(\sum_{i=1}^{n}{|B_{1,i}|},\sum_{i=1}^{n}{|B_{2,i}|},\cdots,\sum_{i=1}^{n}{|B_{n,i}|})$,
where $B_{i,j}$ is the element of $B$. Then $\max
\bar{Q}=\emph{Tr}(S^{T}BS)$ is equivalent to $\max
Q_{m}=\emph{Tr}(S^{T}(B+D)^{m}S)$ for all positive integer number
$m$. $(B+D)^{m}$ is a positive matrix, so the modularity
maximization problem can map to a continuous nonconvex quadratic
programming. These theorems will be detailed in Section $2$. In this
way, modularity maximization problem is equivalent to $\max
Q_{m}=\emph{Tr}(S^{T}(B+D)^{m}S)$. We have done many numerical
experiments on artificial and real-world networks such as
physics-economics scientists cooperation network, E.coli network and
Collage football network, and found that a proper large $m$ is very
helpful for two basic neighborhood transformation algorithms and EO
algorithm for maximizing $Q$. It implies that our results has great
possibility to enhance the efficiency of many heuristic algorithms.

\section{Theorems about modularity maximization problem}

Newman and Givan proposed the modularity $Q$ index based on the
common experience that such networks seem to have communities in
them: subsets of nodes within which node-node connections are dense,
but between which connections are less dense \cite{GN}. According to
\cite{17}, modularity $Q$ can be simplified. Suppose we have a
network $N$ which has $n$ nodes and can be represented
mathematically by an adjacency matrix $A$ with elements $A_{i,j}=1$
if there is an edge from $i$ to $j$ and $A_{i,j}=0$ otherwise.
$d_{i}$ denotes the degree of node $i$ and $P$ is a matrix,
$P_{i,j}=\frac{d_{i}d_{j}}{2L}$. Without losing any generality we
assume that the network $N$ has $n$ communities (if the number of
community is less than $n$ we can use $0$ to substitute). Suppose
$S=(S_{1},S_{2},\cdots,S_{n})$ is the community structure matrix,
$S_{i}\in\{0,1\}^{n}$ denotes the $i$ community, $i=1,2,\cdots,n$.
For example: assume $S_{i}=(0,1,0,1,0,\cdots,0)^{T}$, it denotes
that community $i$ only contains two nodes which are node $2$ and
$4$. Because a node only belongs to one community, each row of $S$
just has one $1$. We use $\mathbb{S}$ to denotes the set of all
possible $S$. Let $B=A-P$, we easily have modularity maximization
problem is equivalent to $\max_{S\in
\mathbb{S}}\,\,\bar{Q}=\emph{Tr}(S^{T}BS)$ \cite{17}, where
\textit{Tr} means \emph{trace} which denotes the sum of diagonal
entries of a matrix.

Now we will map the maximization modularity $Q$ problem to nonconvex
quadratic 0-1 programming. Let $\tilde{S}=\left(\begin{array}{c}
                  S_{1} \\
                  S_{2} \\
                  \vdots \\
                  S_{n}
                \end{array}
\right)$, then $\max_{S\in
\mathbb{S}}\,\,\bar{Q}=\emph{Tr}(S^{T}BS)$ can be write as:

$\max_{S\in \mathbb{S}}\,\, \bar{Q}=\tilde{S}^{T}\left(
\begin{array}{cccc}
B &   &   &   \\
& B &   &   \\
&   & \ddots &   \\
&   &   & B \\
\end{array}
\right)\tilde{S}$
\\$\emph{st}.\left\{
   \begin{array}{l}
 s_{1,1}+s_{1,2}+\cdots+s_{1,n}=1 \\
 s_{2,1}+s_{2,2}+\cdots+s_{2,n}=1 \\
 \cdots   \cdots   \cdots \\
 s_{n,1}+s_{n,2}+\cdots+s_{n,n}=1\\
 s_{i,j}\in\{0,1\}\,\,\,i,j=1,2,\cdots,n
   \end{array}
 \right.$

From the subject conditions we can easily get that the set
$\mathbb{S}$ contain $n^{n}$ elements,
$\mathbb{S}=\{S^{1},S^{2},\cdots,S^{n^{n}}\}$. According to the
definition of $\tilde{S}$ we also have the corresponding set
$\tilde{\mathbb{S}}=\{\tilde{S}^{1},\tilde{S}^{2},\cdots,\tilde{S}^{n^{n}}\}$.

\textbf{Theorem\,1}: Let
$D=diag(\sum_{i=1}^{n}{|B_{1,i}|},\sum_{i=1}^{n}{|B_{2,i}|},\cdots,\sum_{i=1}^{n}{|B_{n,i}|})$,
then, $\max_{S\in \mathbb{S}}\,\,\bar{Q}=\emph{Tr}(S^{T}BS)$ problem
is equivalent to the maximization problem of $\max_{S\in
\mathbb{S}}\,\,Q_{1}=\emph{Tr}(S^{T}(B+D)S)$ which can be map to a
nonconvex  quadratic continuous programming.

\textbf{Proof}:

$\because\ \
Tr(S^{T}(B+D)S)=Tr(S^{T}BS)+Tr(S^{T}DS)=Tr(S^{T}BS)+\sum_{i=1}^{n}D_{i,i}$.

$\therefore\ \ \max_{S\in \mathbb{S}}\,\,Q=\emph{Tr}(S^{T}BS)$
problem is equivalent to the maximization problem of

$\max_{S\in \mathbb{S}}\,\,Q=\emph{Tr}(S^{T}(B+D)S)$.

According to \emph{Gerschgorin Circle Theory}
\cite{Numerical_Algbra}, easily we have $B+D$ is a symmetrical
positive matrix.

$\max_{S\in \mathbb{S}}\,\,Q=\emph{Tr}(S^{T}(B+D)S)$ is a continuous
nonconvex quadratic programming \cite{Globle Optimization}.

\textbf{Theorem\,2}: For all positive integer number $m$,
$\max_{S\in \mathbb{S}}\,\,\bar{Q}=\emph{Tr}(S^{T}BS)$ problem is
equivalent to the maximization problem of $\max_{S\in
\mathbb{S}}\,\,Q_{m}=\emph{Tr}(S^{T}(B+D)^{m}S)$.

\textbf{Proof}: $\because\ \ \max_{S\in
\mathbb{S}}\,\,\bar{Q}=\emph{Tr}(S^{T}BS)$

is equivalent to $\max_{S\in
\mathbb{S}}\,\,Q_{1}=\emph{Tr}(S^{T}(B+D)S)$

is equivalent to $\max_{\tilde{S}\in \tilde{\mathbb{S}}}\,\,
Q_{1}=\tilde{S}^{T}\left(
                            \begin{array}{cccc}
                              B+D &   &   &   \\
                                & B+D &   &   \\
                                &   & \ddots &   \\
                                &   &   & B+D \\
                            \end{array}
                          \right)\tilde{S}$

and $\tilde{S}^{T}\tilde{S}=Tr(S^{T}S)=n$

$\therefore\ \ \max_{\tilde{S}\in
\tilde{\mathbb{S}}}\,\,Q_{1}=\tilde{S}^{T}\left(
                            \begin{array}{cccc}
                              B+D &   &   &   \\
                                & B+D &   &   \\
                                &   & \ddots &   \\
                                &   &   & B+D \\
                            \end{array}
                          \right)\tilde{S}$

is equivalent to  $\max_{\tilde{S}\in
\tilde{\mathbb{S}}}\,\,Q_{m}=\tilde{S}^{T}\left(
                            \begin{array}{cccc}
                              B+D &   &   &   \\
                                & B+D &   &   \\
                                &   & \ddots &   \\
                                &   &   & B+D \\
                            \end{array}
                          \right)^{m}\tilde{S}$

is equivalent to  $\max_{\tilde{S}\in
\tilde{\mathbb{S}}}\,\,Q_{m}=\tilde{S}^{T}\left(
                            \begin{array}{cccc}
                              (B+D)^m &   &   &   \\
                                & (B+D)^m &   &   \\
                                &   & \ddots &   \\
                                &   &   & (B+D)^m \\
                            \end{array}
                          \right)\tilde{S}$

is equivalent to $\max_{S\in
\mathbb{S}}\,\,Q_{m}=\emph{Tr}(S^{T}(B+D)^{m}S)$.

\section{Application of the theorems}

\begin{figure}
\center
\includegraphics[height=4cm,width=6cm]{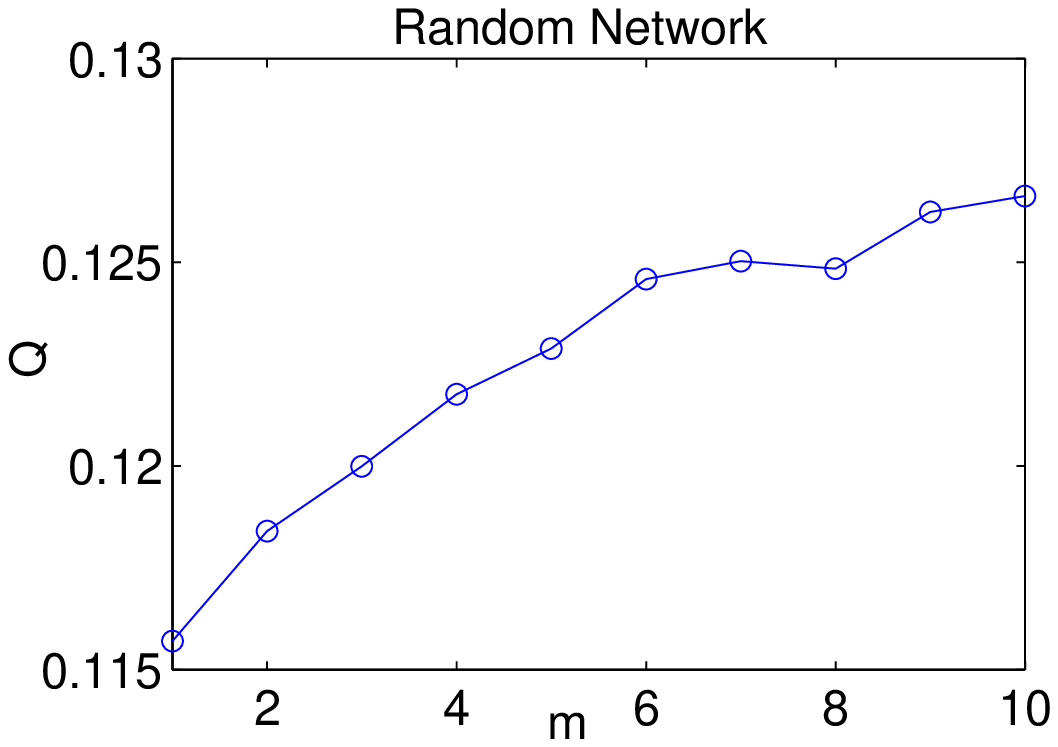}
\includegraphics[height=4cm,width=6cm]{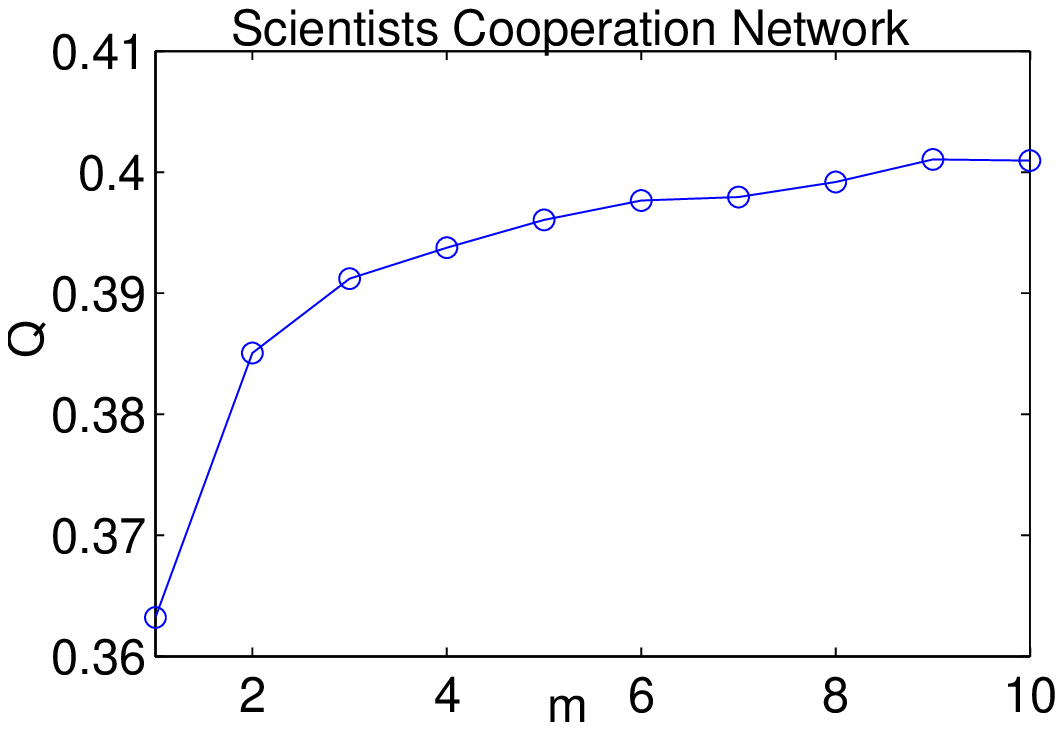}
\includegraphics[height=4cm,width=6cm]{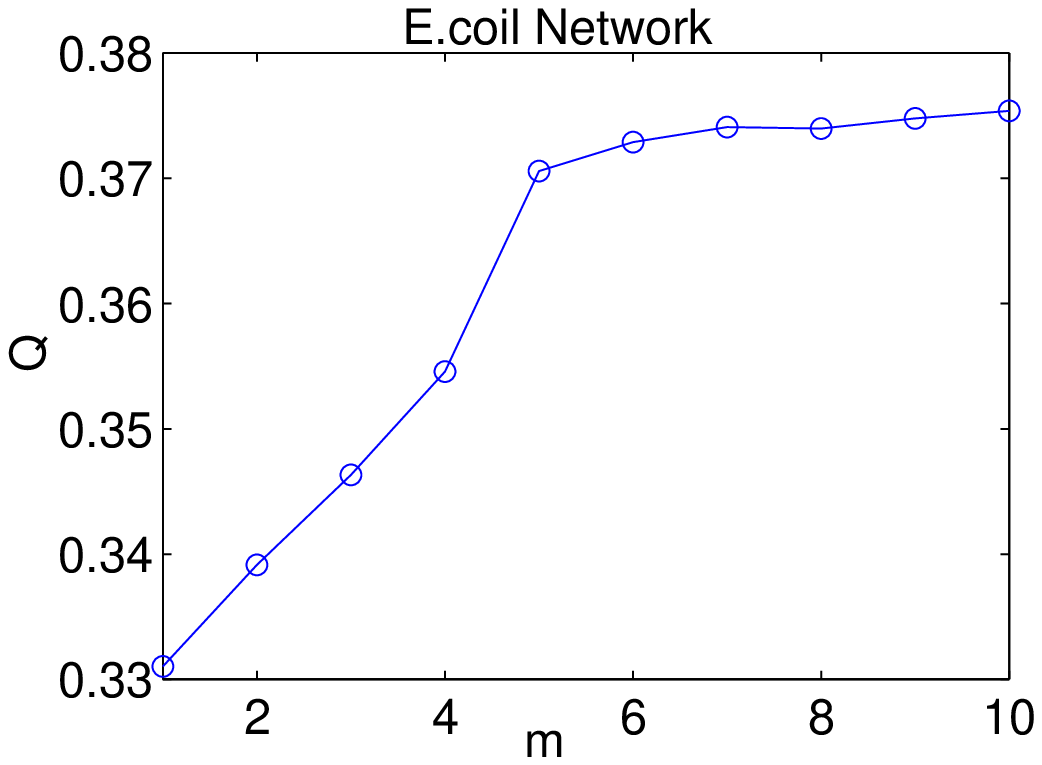}
\includegraphics[height=4cm,width=6cm]{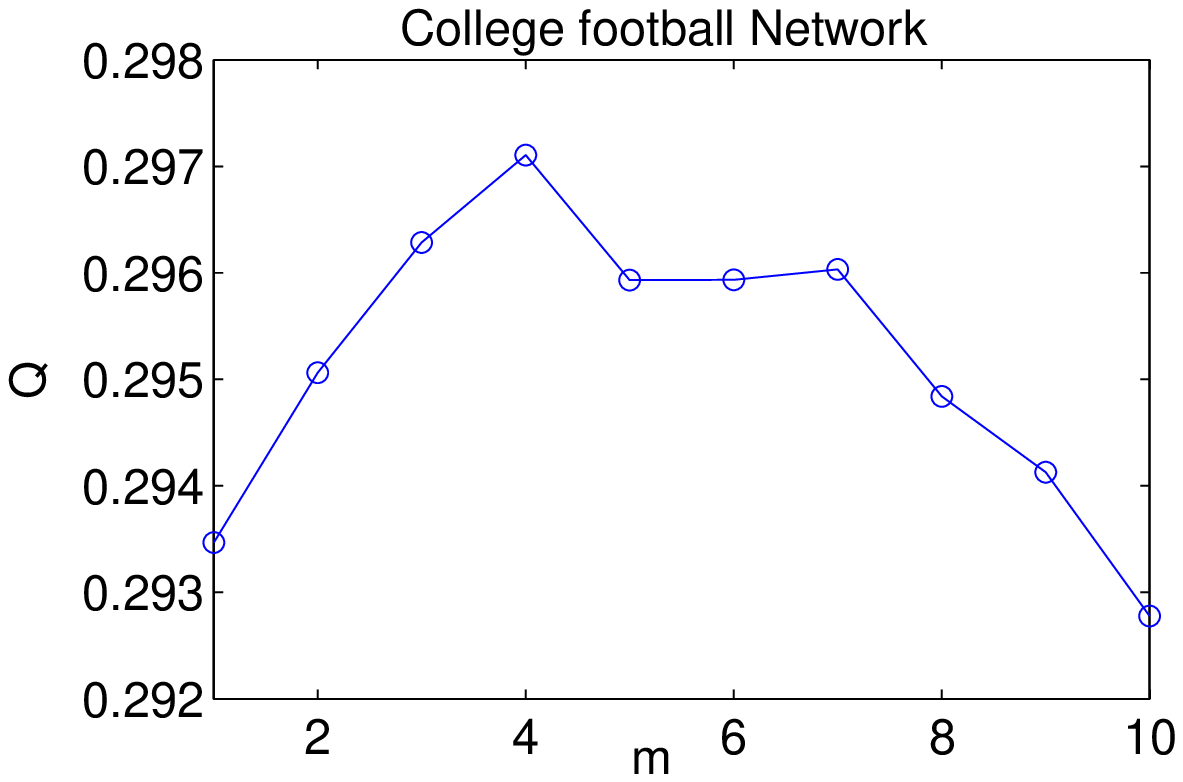}
\caption{Results of random neighborhood transformation algorithm
.}\label{expriments2}
\end{figure}

\begin{figure}
\center
\includegraphics[height=4cm,width=6cm]{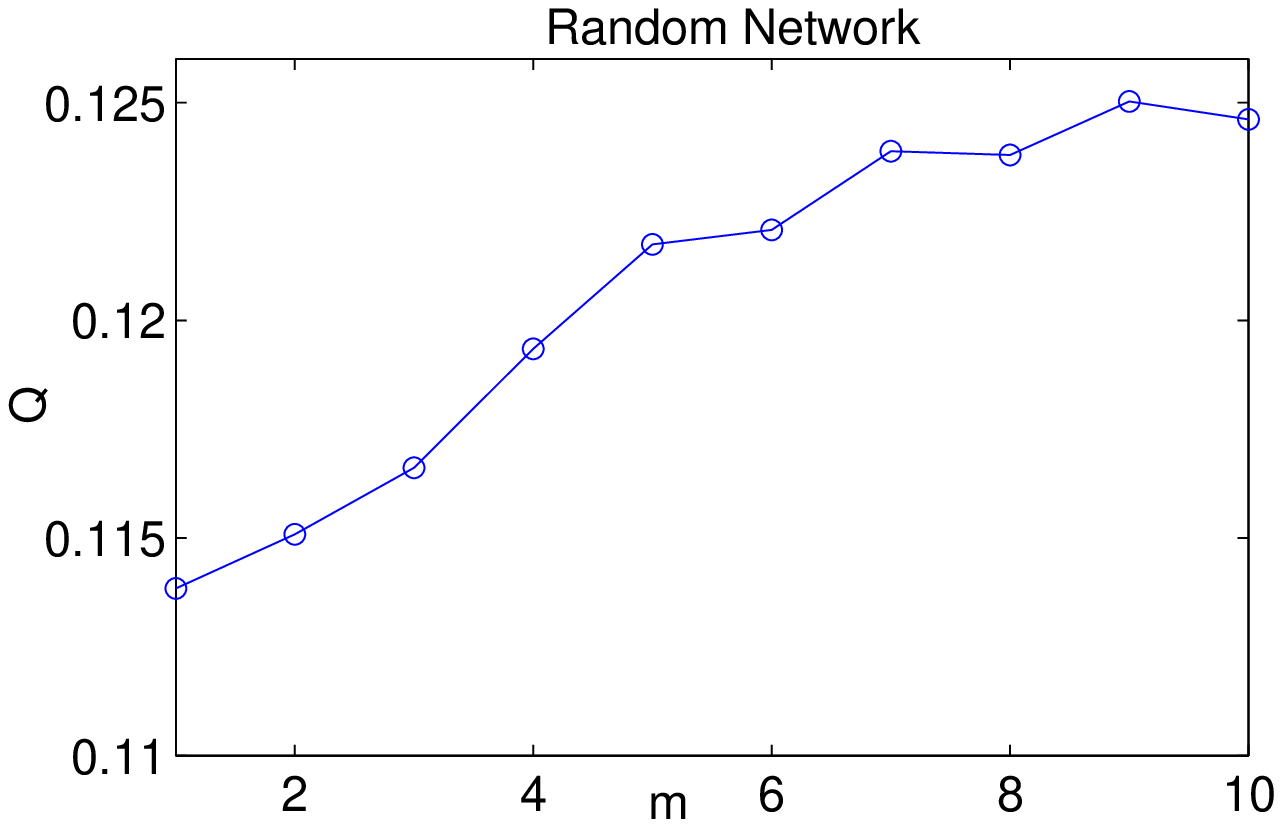}
\includegraphics[height=4cm,width=6cm]{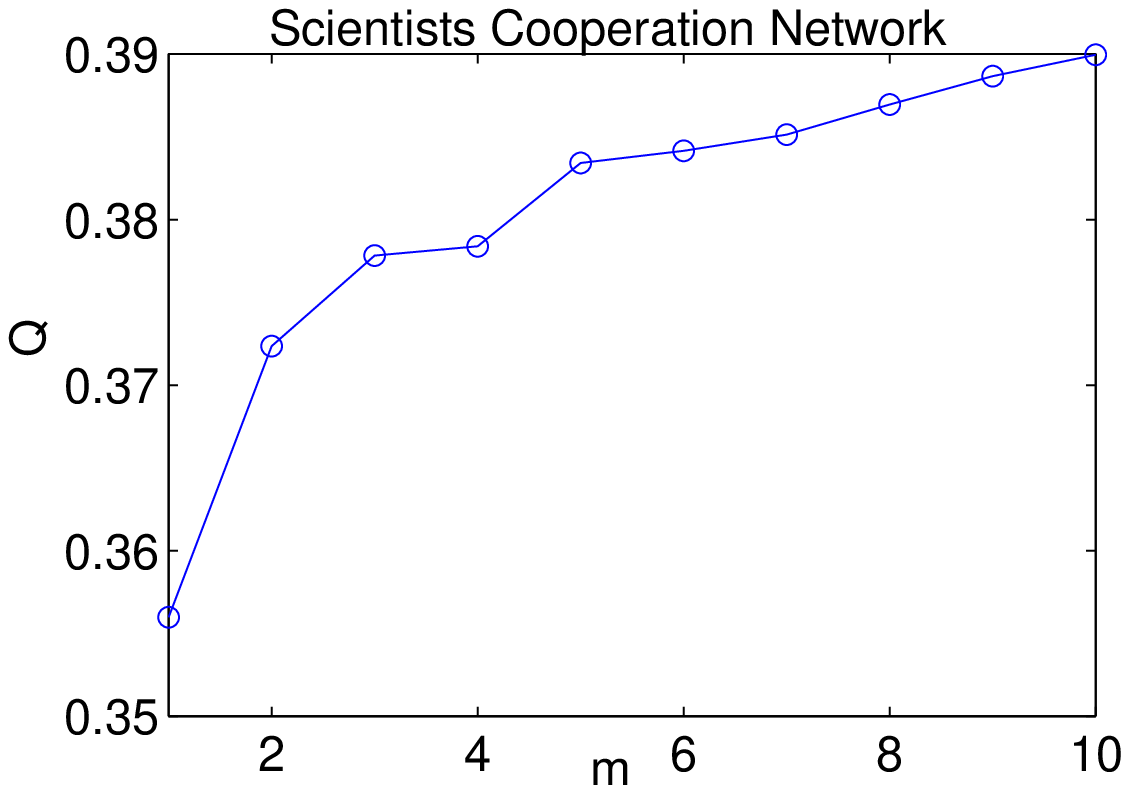}
\includegraphics[height=4cm,width=6cm]{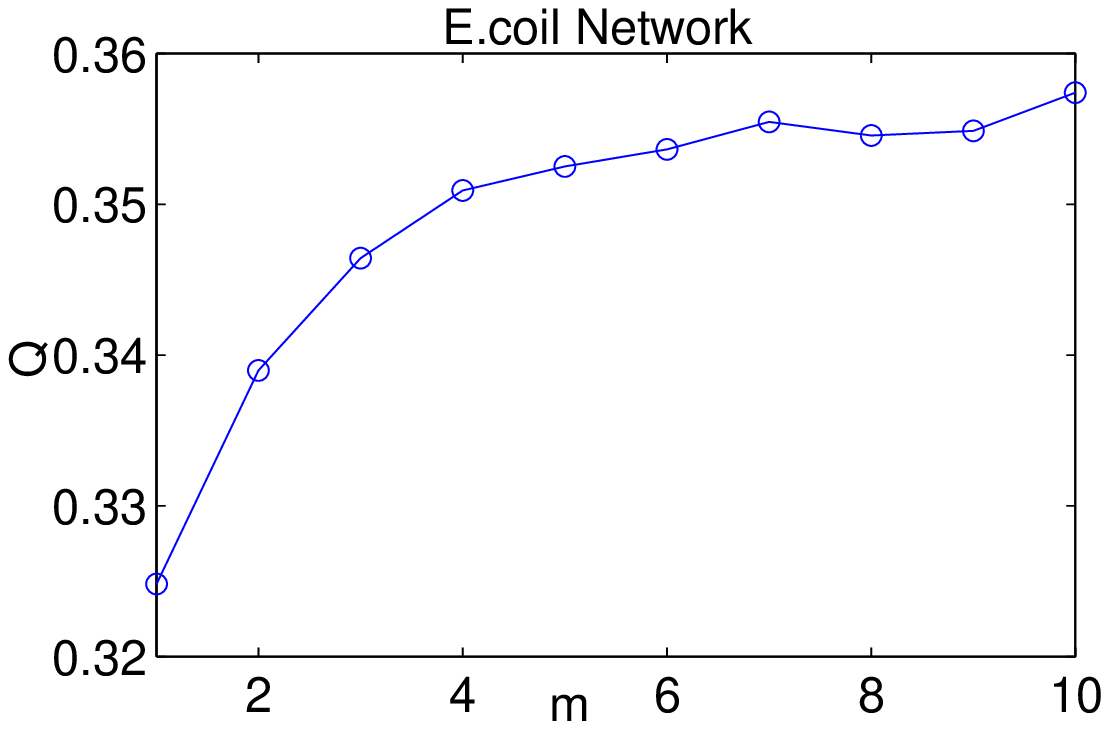}
\includegraphics[height=4cm,width=6cm]{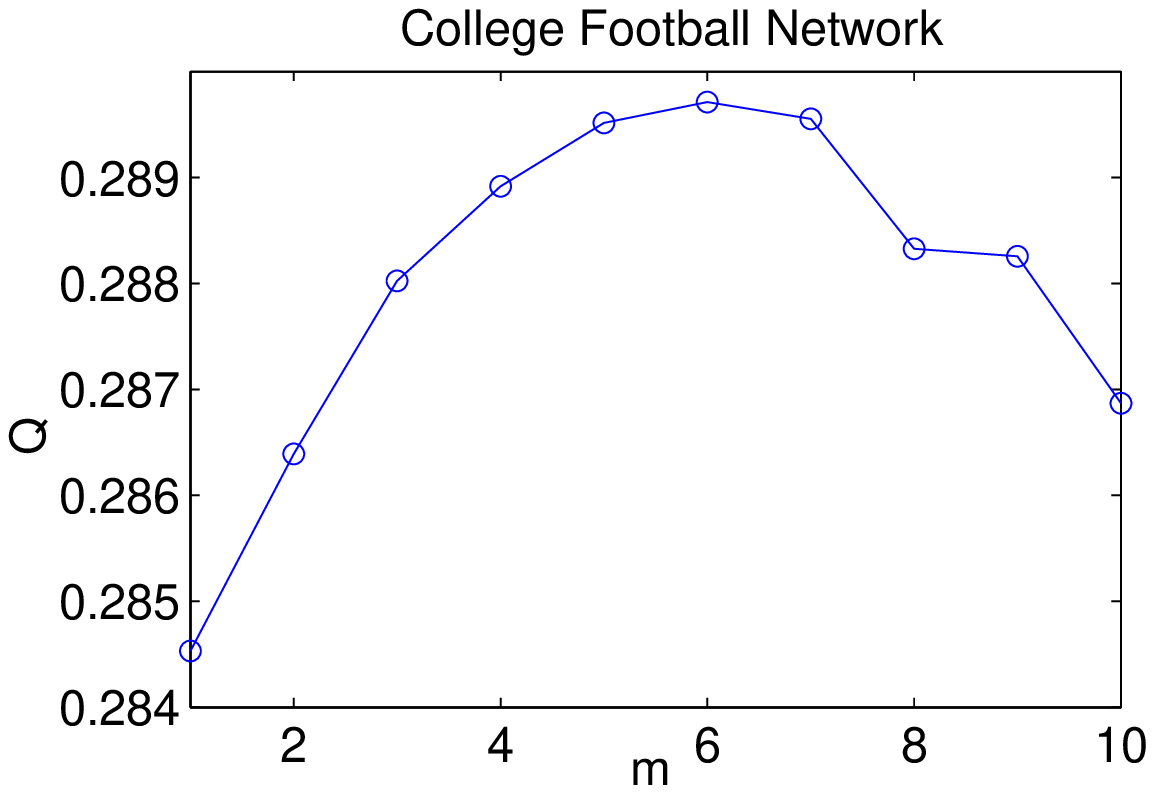}
\caption{Results of greedy neighborhood transformation
algorithm.}\label{expriments3}
\end{figure}

Based on the theorem 2, maximizing $Q$ is equivalent to $\max_{S\in
\mathbb{S}}\,\,Q_{m}=\emph{Tr}(S^{T}(B+D)^{m}S)$. Can we enhance the
efficiency of heuristic algorithms for maximizing modularity $Q$ by
changing it into this new maximizing problem with a proper large
$m$? There are so many heuristic algorithm for maximizing modularity
$Q$, we cannot investigate all of them. If we could, we also cannot
promise our method satisfy the future heuristic algorithms. But it
is well-know that, for many heuristic algorithms such as EO, Potts
\cite{Potts} and so on, their key methods are to find optimal
neighborhood transformations, where neighborhood transformation
means moving a node for one community to another community at each
optimizing step. So if our method is effective on the basic
neighborhood transformation algorithms, it will has great
possibility to be effective on many other heuristic algorithms.
There are two basic neighborhood transformation algorithms. One is
random neighborhood transformation algorithm. We randomly initiate
the beginning partition (with sufficient number of groups), then at
each step, randomly choose a node form one community and move it
into another one that can make $Q$ become larger, until moving any
node cannot make $Q$ larger any more. The other algorithm is greedy
neighborhood transformation algorithm. The corresponding process is
similar with the process of random one, but the difference is that
at each step, the node will be moved to a group that makes $Q$ has
the largest increment. We choose four different fields' networks to
test our method. One is the classical artificial random network
which has $n=128$ nodes divided into $4$ communities of $32$ nodes
each. Edges between two nodes are introduced with different
probabilities depending on whether the two nodes belong to the same
community or not: every node has $<k_{intra}=8>$ links on average to
its fellows in the same community, and $<k_{inter}=8>$ links to the
outer-world. Here we chose the artificial network with the diffuse
community structures to test our method. It is because when the
network contains clear community structure, $m$ has almost no
effects on the final partition. The rest $3$ networks are scientists
cooperation network \cite{Econophysicist}, E.coli network
\cite{Barabasi www data} and college football network \cite{GN}. The
results show that for a proper large $m$, our method is helpful for
finding large value of $Q$ (as shown in Fig. \ref{expriments2} and
Fig. \ref{expriments3}). But it is hard to say it need more or less
time in maximizing $Q$ process.

We also use the extremal optimization algorithm (EO) \cite{EO} to
test our method. EO was proposed by Jordi Duch and Alex Arenas,
which is heuristic algorithm. In their algorithm, they define a
fitness of each node. The fitness $f_{i}$ of node $i$ is defined as
\begin{equation}f_{i}=\frac{q_{i}}{k_{i}}\label{fitness}\end{equation}
where, $k_{i}$ denotes the degree of node $i$, and the $q_{i}$ is
the contribution of individual node $i$ to the $Q$. Assume $e_{i}$
denotes the $n$-dimensional vector in which the $i$th element is
$1$, others $0$ then
\begin{equation}q_{i}=e_{i}^{T}BS_{i}\label{contribution}\end{equation}
For the maximization problem $\max_{S\in
\mathbb{S}}\,\,Q=\emph{Tr}(S^{T}(B+D)^{m}S)$, the contribution is
\begin{equation}q_{i}^{m}=e_{i}^{T}(B+D)^{m}S_{i}\label{contribution2}\end{equation}
Unfortunately, we cannot use the function
$f_{i}^{m}=\frac{q_{i}^{m}}{k_{i}}$ (as Eq. \ref{fitness}) to define
the fitness, for it is not satisfy the original conditions (see
\cite{EO}). So we define the new fitness function as the Eq.
\ref{contribution2}. Moreover, Jordi Duch and Alex Arenas didn't
define the `optimal state' quantitatively in \cite{EO}. In this
paper, we think a partition process has arrived the optimal state at
step $t$ if the $Q$ of $t$ is equal or larger than each $Q$ from
step $t+1$ to $t+n$, where $n$ is the node number of a network.

\begin{figure}
\center
\includegraphics[height=4cm,width=6cm]{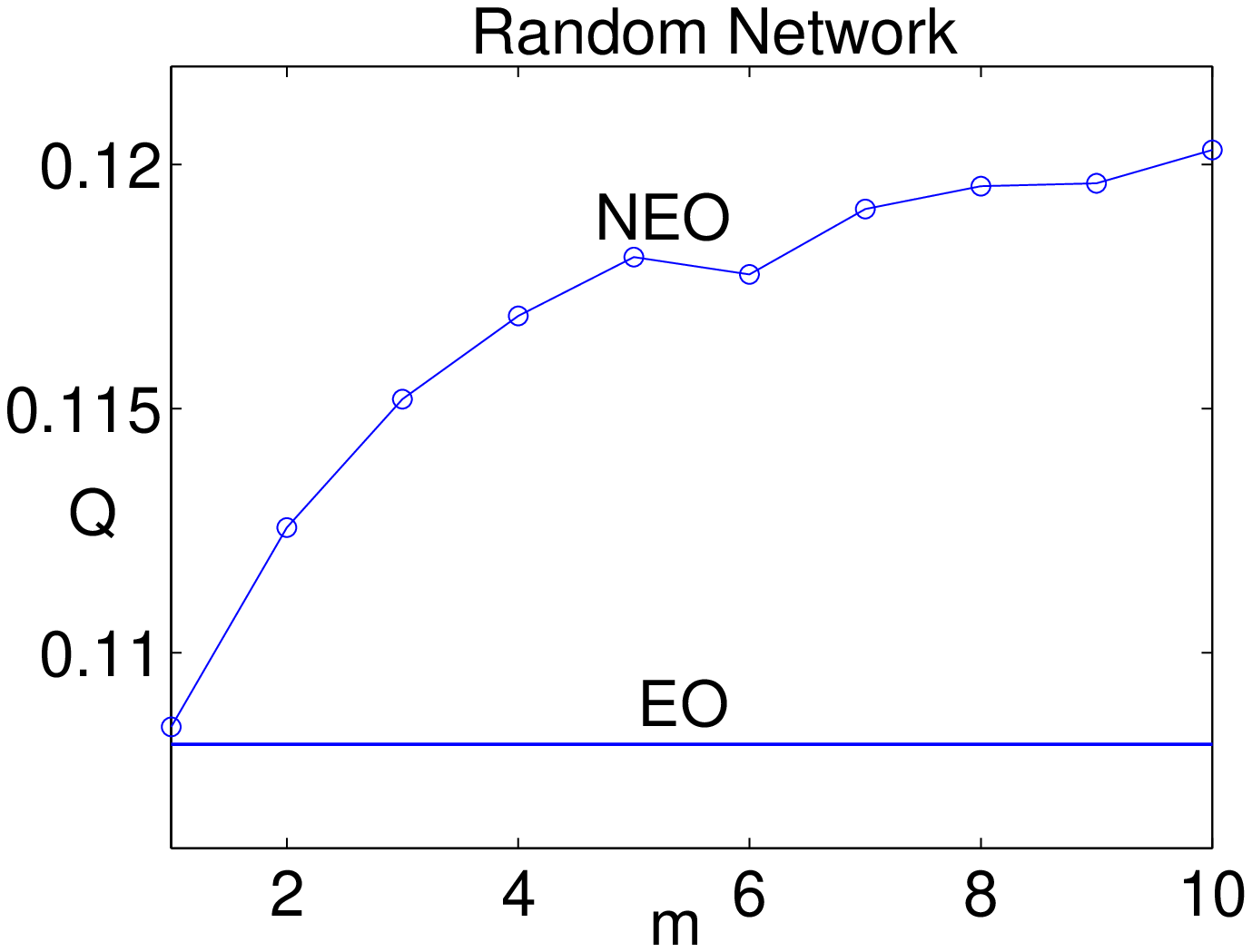}\includegraphics[height=4cm,width=6cm]{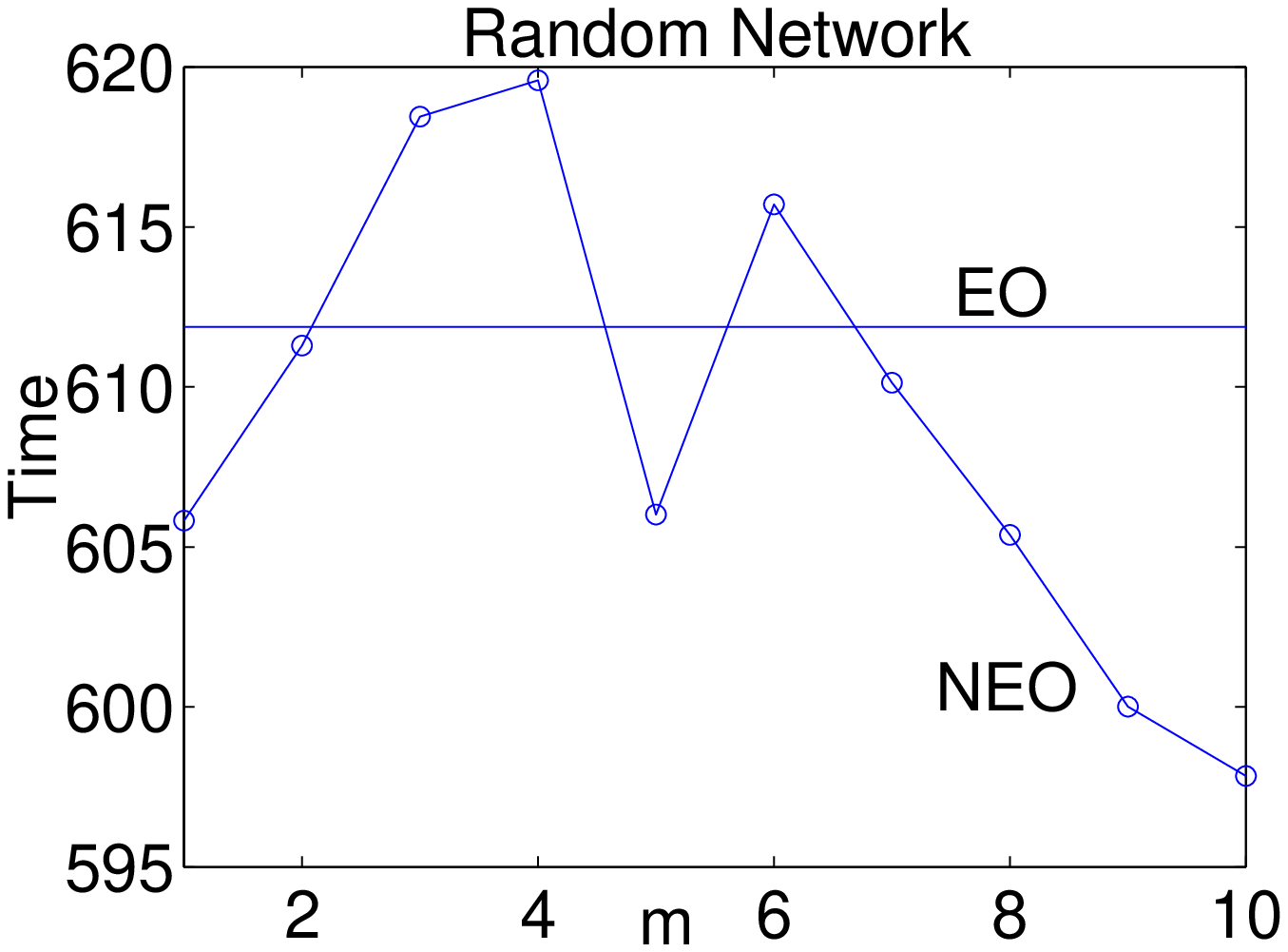}
\includegraphics[height=4cm,width=6cm]{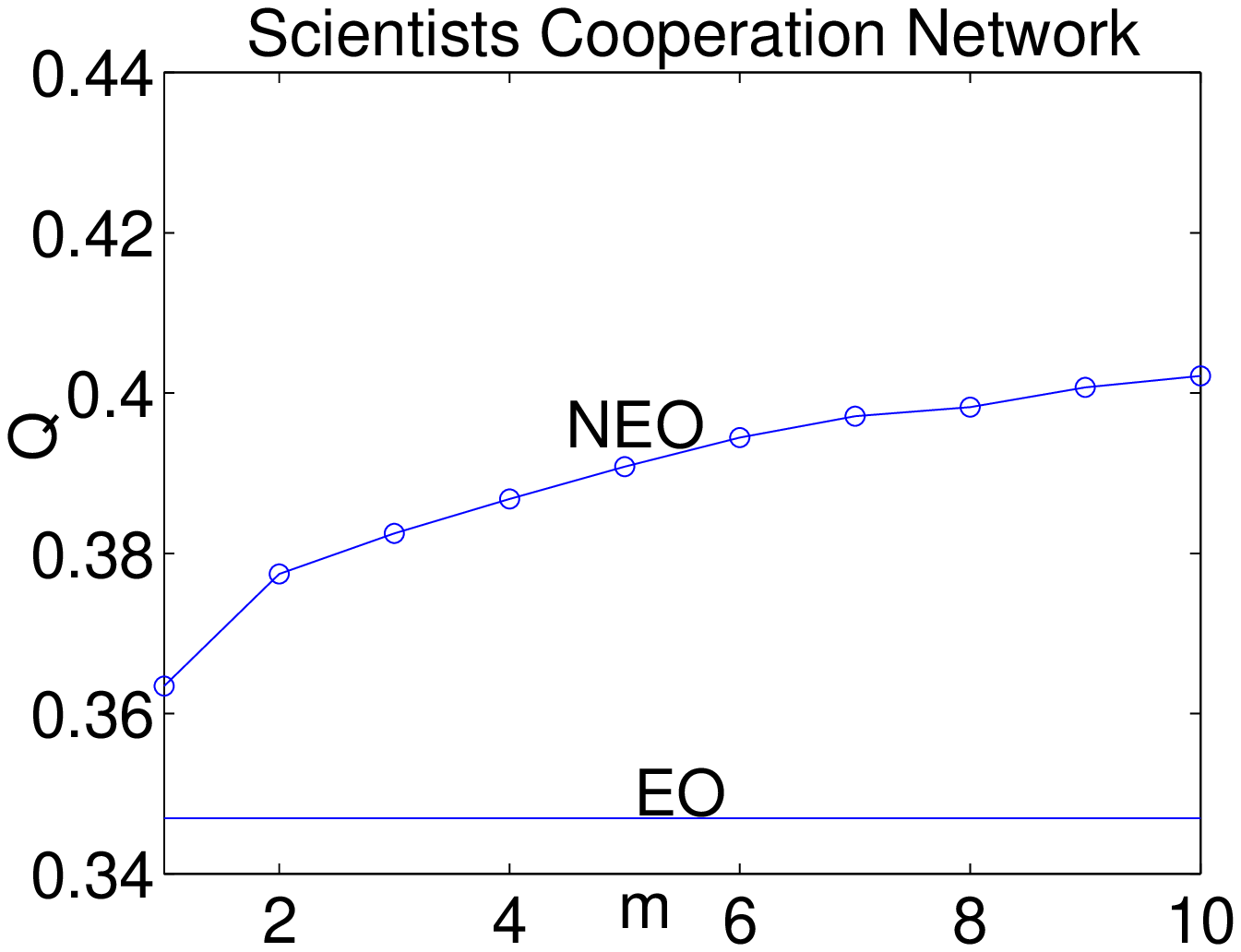}\includegraphics[height=4cm,width=6cm]{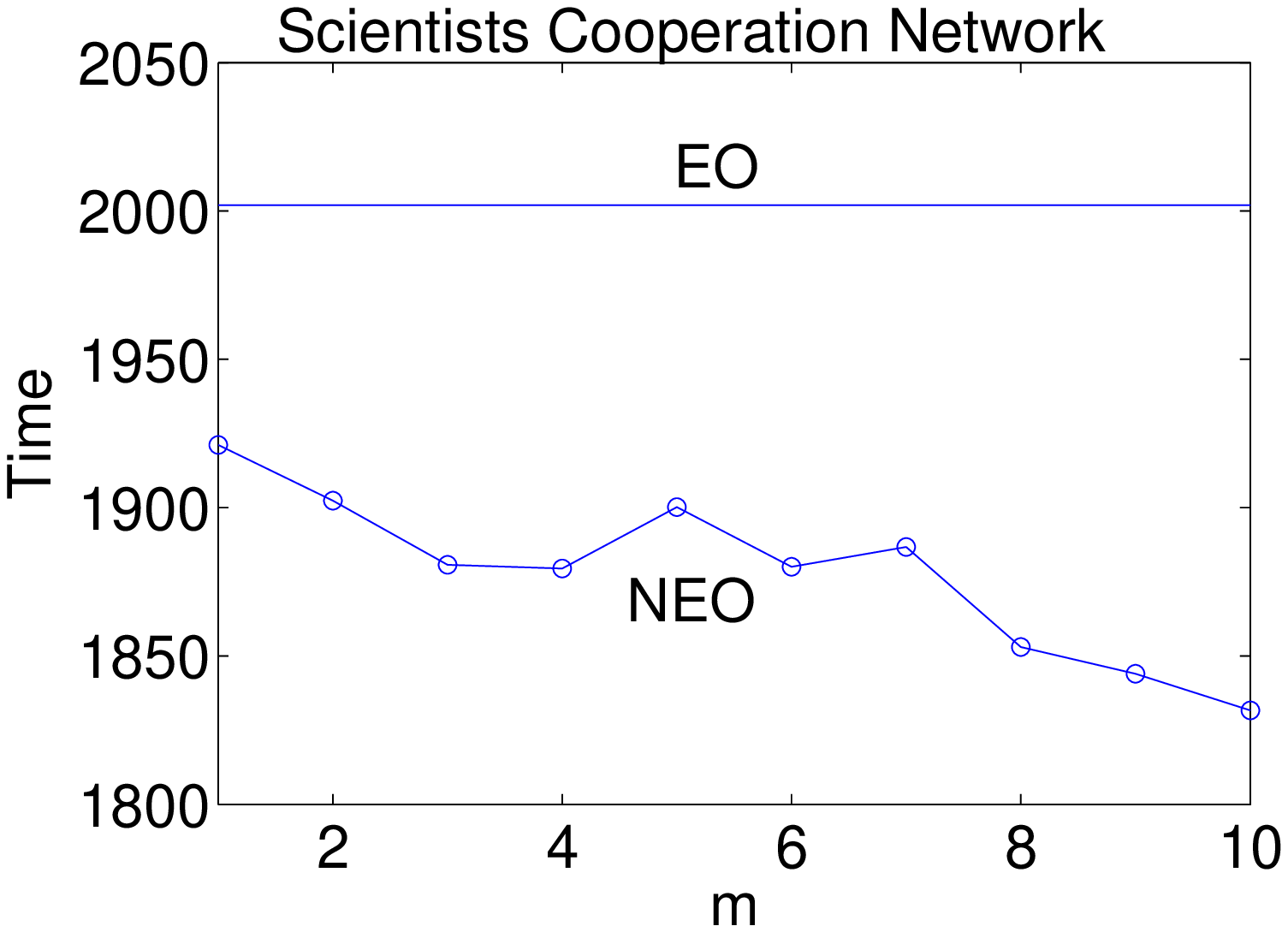}
\includegraphics[height=4cm,width=6cm]{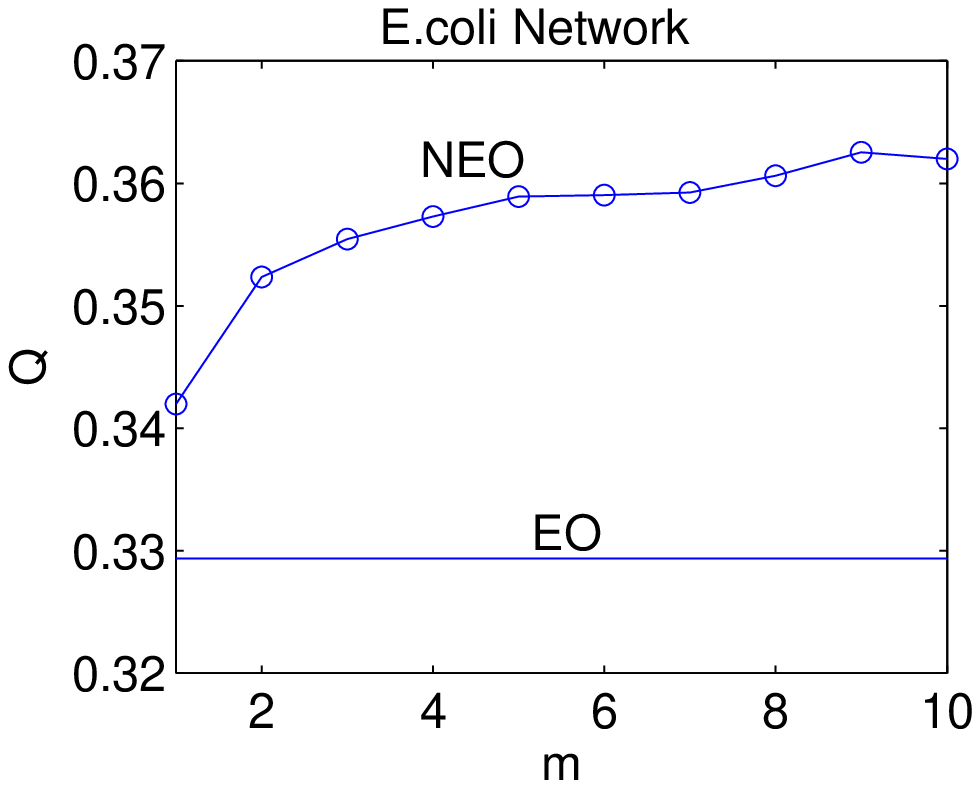}\includegraphics[height=4cm,width=6cm]{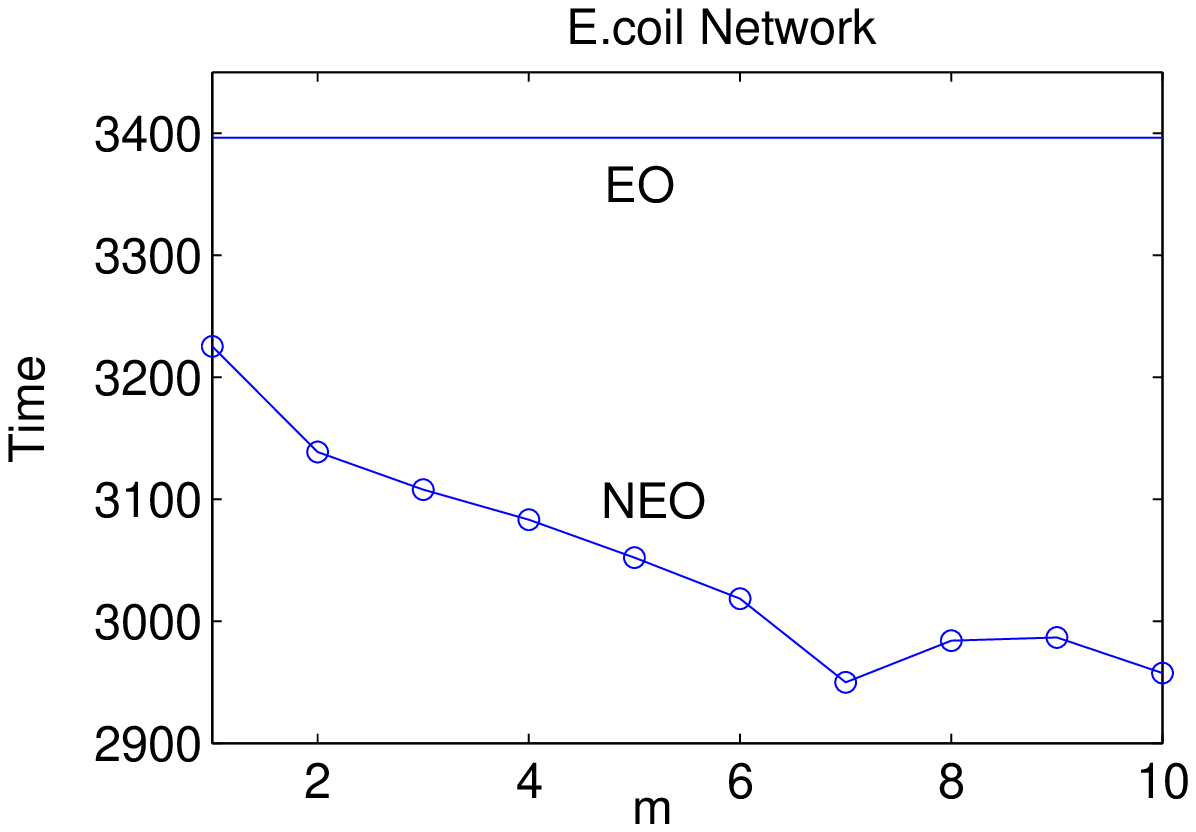}
\includegraphics[height=4cm,width=6cm]{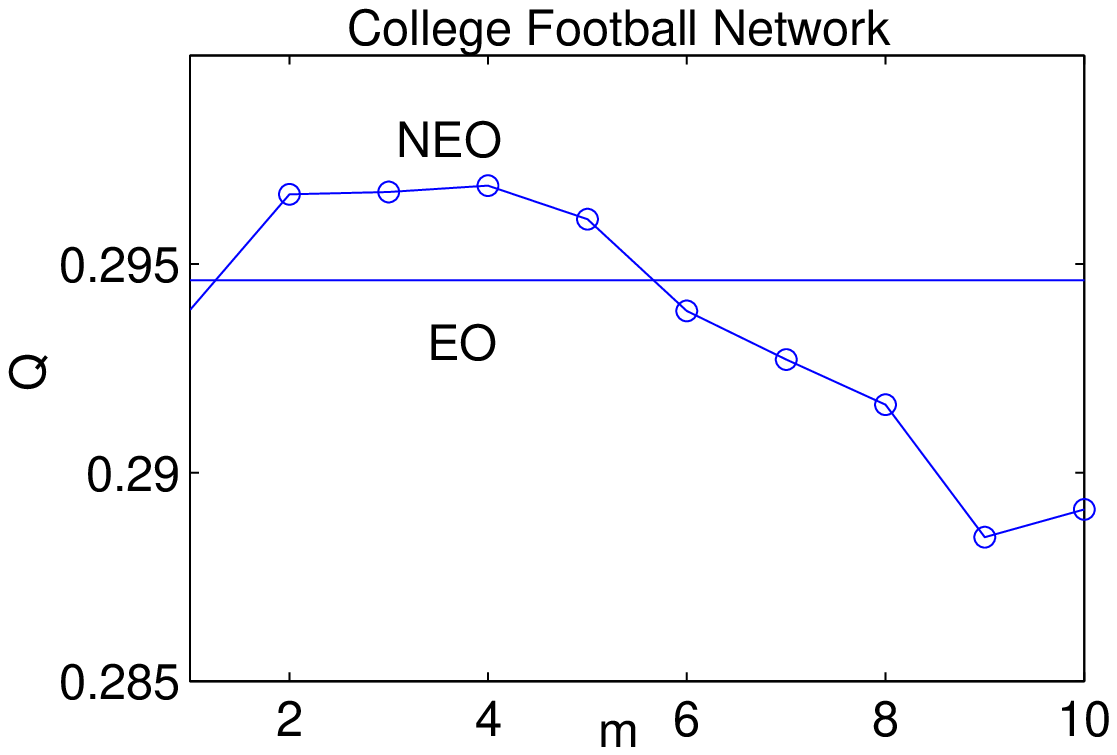}\includegraphics[height=4cm,width=6cm]{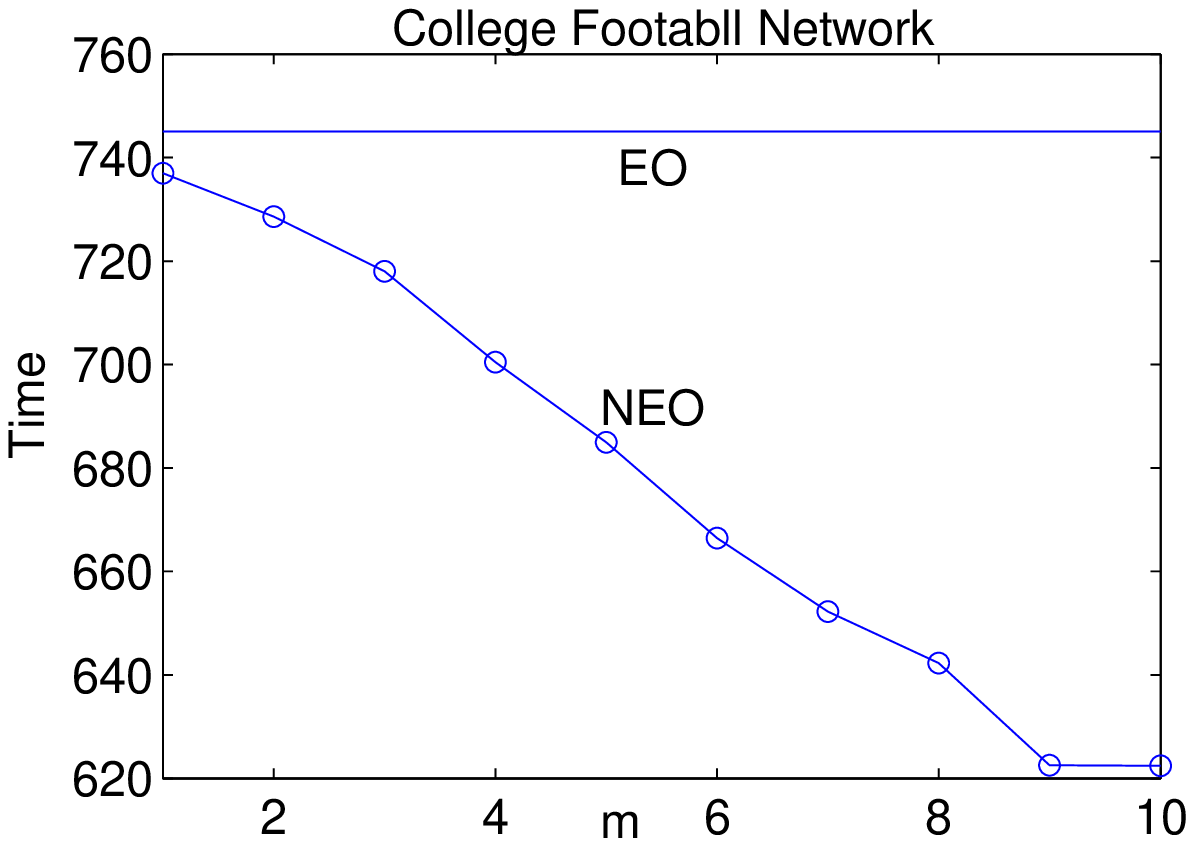}
\caption{These plots show the results of NEO and EO algorithm in
different networks. The line means the $Q$ value which is got by EO
(with original fitness function). From these plots we can conclude
that large $m$ is very helpful in both maximizing $Q$ and reducing
time complicity. But sometimes too large $m$ will bring overdone
effects such as the results in the collage football
network.}\label{expriments}
\end{figure}

We investigate extremal optimization with new fitness function (NEO)
for different $m$ and compare the NEO algorithm with the EO
algorithm in the above four networks. The results show that the
proper larger $m$ is very helpful both for maximizing modularity $Q$
and reducing computing time, but sometimes the too large $m$ is not
helpful (as shown in Fig. \ref{expriments}). We guess one of the
main reasons is that too large $m$ will bring more computing errors.

\section{Conclusion and discussion}
We prove that the modularity maximization problem is equivalent to a
nonconvex quadratic programming problem. Based the characteristics
of nonconvex quadratic programming, we demonstrate that the
modularity maximization problem is equivalent to the maximization
problem $\max_{S\in
\mathbb{S}}\,\,Q_{m}=\emph{Tr}(S^{T}(B+D)^{m}S)$. This conclusion
provide a simple way to improve the efficiency of algorithms for
maximizing modularity $Q$. Many numerical experiments are done in
different networks include artificial networks, scientists
cooperation network, E.coli network and Collage football network.
The results show that new maximization problem with proper large $m$
can enhance the efficiency of the heuristic algorithms for
maximizing $Q$. Especially, it is helpful in both maximization $Q$
and time complicity for EO algorithm. But it is a real challenge
problem to strictly give the most optimal $m$.

\section*{Acknowledgement}
The authors want to thank M. E. J. Newman from providing the college
football network and Qiang Yuan for some useful discussion. This
work is partially supported by 985 Projet and NSFC under the grant
No.$70431002$, No.$70771011$.

\end{document}